\documentclass[12pt,preprint]{aastex}

\shorttitle{Occultation Spectroscopy of the Secondary Eclipse of HD~209458b}
\shortauthors{Richardson et al.}

\newcommand{\tp}{HD~209458b}
\newcommand{\tar}{HD~209458}
\newcommand{\comp}{HD~210483}

\begin{document}

\title{Infrared Observations During the Secondary Eclipse of HD~209458b \\
 I. 3.6-Micron Occultation Spectroscopy Using the VLT \footnote{Based
 on observations collected at the European Southern Observatory, Chile (proposal no. 67.C-0196)}}


\author{L. Jeremy Richardson\altaffilmark{2,3,4}, Drake
Deming\altaffilmark{2}, Guenter Wiedemann\altaffilmark{5}, Cedric Goukenleuque\altaffilmark{2,6}, David Steyert\altaffilmark{2,6,7}, Joseph Harrington\altaffilmark{8}, and Larry W. Esposito\altaffilmark{3}}

\email{lee.richardson@colorado.edu}
\email{ddeming@pop600.gsfc.nasa.gov}
\email{gwiedema@astro.uni-jena.de}
\email{cedric@pseudonomen.gsfc.nasa.gov}
\email{steyert@lepvax.gsfc.nasa.gov}
\email{jh@oobleck.astro.cornell.edu}
\email{larry.esposito@lasp.colorado.edu}

\altaffiltext{2}{Planetary Systems Branch, Code 693, Goddard Space Flight Center, Greenbelt, MD 20771}
\altaffiltext{3}{Laboratory for Atmospheric and Space Physics,
University of Colorado, 1234 Innovation Drive, Boulder,~CO~80303}
\altaffiltext{4}{Department of Physics, University of Colorado,
Boulder, CO 80309}
\altaffiltext{5}{Astrophysikalisches Institut und
Universitats-Sternwarte, D-07745 Jena, Germany}
\altaffiltext{6}{NAS/NRC Research Associate}
\altaffiltext{7}{Joint Center for Earth Systems Technology, University
 of Maryland Baltimore County}
\altaffiltext{8}{Department of Astronomy, Cornell University, Ithaca, NY 14853-6801}

\begin{abstract}
We search for an infrared signature of the transiting extrasolar
planet \tp\ during secondary eclipse.  Our method, which we
call `occultation spectroscopy,' searches for the disappearance and
reappearance of weak spectral features due to the exoplanet as it passes
behind the star and later reappears.  We argue that at the longest
infrared wavelengths, this technique becomes preferable to
conventional `transit spectroscopy'.  We observed the system in the wing of the
strong $\nu_3$ band of methane near $3.6$ \micron\ during two secondary
eclipses, using the VLT/ISAAC spectrometer at a spectral resolution of
3300.  Our analysis, which utilizes a model template spectrum,
achieves sufficient precision to expect detection
of the spectral structure predicted by an irradiated, low-opacity (cloudless),
low-albedo, thermochemical equilibrium model for the exoplanet
atmosphere. However, our observations show no evidence for the
presence of this spectrum from the exoplanet, with the statistical
significance of the non-detection depending on the timing of the
secondary eclipse, which depends on the assumed value for the orbital
eccentricity.  Our results reject certain specific models of the atmosphere of
\tp\ as inconsistent with our observations at the $3\sigma$ level,
given assumptions about the stellar and planetary parameters.
\end{abstract}

\bigskip
\bigskip
\bigskip
\bigskip



\section{INTRODUCTION}

The discovery of the transiting extrasolar planet \tp\
\citep{cb00,hmbv00} has provided a unique opportunity to deduce the
physical characteristics of one example of the so-called `hot Jupiter'
class of exoplanets.
The transit geometry has allowed an accurate derivation of the stellar and
planetary mass and radius \citep{bc01} and also provides an
opportunity to measure the composition of the exoplanet atmosphere.  The scale
height of an atmosphere in hydrostatic equilibrium is proportional to
$T/g$, where $T$ is the atmospheric temperature and $g$ is the surface
gravity. Because stellar irradiation of the planet is intense at
$0.046$~AU \citep{ss98}, the atmosphere could be as hot as
$T \sim 1400$~K or more.  Coupled with the relatively low surface gravity ($g =
848$~cm~s$^{-2}$), the atmospheric scale height is potentially as large
as $H \sim 750$~km. Moreover, when viewed tangent to the limb, the
atmosphere can be opaque over several scale heights at the wavelength
of a strong absorption line.  The effective absorbing area of the planet
can thus be greater in a strong line by a detectable amount, producing
a slightly deeper eclipse at wavelengths close to the line core
\citep{ss00,brown01,hubbard}. That is, the opacity of the exoplanet atmosphere
will impose weak, but potentially detectable, absorption lines on the
stellar spectrum as the planet passes in front of the star.  Several
investigators have attempted to detect this effect
\citep{bundy,moutou01}.  Using the
Hubble Space Telescope to observe the primary eclipse, \citet{cb02}
succeeded in detecting the sodium doublet in the atmosphere of \tp.

In addition to sodium, absorption in molecular features of water,
methane, and carbon monoxide may also be observable
\citep{hubbard, brown01}. \citet{bc02} attempted to detect absorption
by carbon monoxide using ground-based observations. From their attempt
and our own experience, we recognize that the terrestrial atmosphere can
potentially frustrate such efforts, but we believe that good observing
conditions and appropriate observational and analytical techniques
will enable successful measurements from the ground.  The known timing of the
transit \citep{bc01} is a boon to this technique, because it
discriminates against false signals by providing a temporal modulation
on the signal.

In this paper we introduce a new observational approach involving the
{\it secondary eclipse}, when the planet passes
behind the star.  The importance of the secondary eclipse was first
pointed out by \citet{cb00}, but no measurement attempt has been
reported in the literature.  We have performed moderate resolution
`occultation spectroscopy' to measure the modulation (i.e.,
disappearance and reappearance) of the potential exoplanet line
spectrum as the planet is hidden and revealed by the star.  We have
also performed lower resolution spectroscopy during and near secondary
eclipse using the NASA Infrared Telescope Facility (IRTF), attempting
to isolate the broad infrared (IR) flux peaks from
the planet's presumably hot atmosphere at wavelengths where the
opacity is minimal.  This paper introduces the concept of occultation
spectroscopy for extrasolar planets and reports the Very Large
Telescope (VLT) spectroscopic results; lower
resolution occultation spectroscopy of the IR flux peaks using the
IRTF observations will be the subject of a second paper.

\section{OCCULTATION SPECTROSCOPY}

In this section, we briefly explain our rationale for attempting
occultation spectroscopy at secondary eclipse, as opposed to more
conventional transmission spectroscopy during primary eclipse.  At
visible and near-IR wavelengths, the self-emitted flux of the
planet is negligible, making occultation spectroscopy of thermal emission
impossible.  However,
many of the strongest molecular features of interest in exoplanets
(e.g., the very strong fundamental $\nu_3$ band of methane) occur at
longer IR wavelengths.  With increasing wavelength, the total signal
detectable using occultation spectroscopy becomes as large, or larger,
than that from transit spectroscopy.  To see this, consider a spectral line
having sufficient strength to be opaque $N$ scale heights higher in
the atmosphere compared to the nearby continuum when
viewed tangent to the limb.  Cases of interest are for large $N$, so
we can safely assume that the line is also optically thick when viewed
at normal incidence to the exoplanet atmosphere.  Transit spectroscopy
looks for this absorption against a stellar continuum flux
proportional $\sim 2\pi RNH B^{*}_\lambda$, where $H$ is the
atmospheric scale height of the exoplanet, $ 2\pi RNH$ is the area of
the annulus of the planetary atmosphere, and
$B^{*}_\lambda$ is the
Planck function for the stellar continuum temperature at wavelength
$\lambda$.  Occultation spectroscopy, on the other hand, looks for the
modulation of the
exoplanet spectrum itself, with continuum flux proportional to $\sim
\pi R^{2} B^{p}_\lambda$, where $B^{p}_\lambda$ is the Planck function
for the exoplanet continuum temperature.  With increasing wavelength,
the increase in $B^{p}_\lambda$ and the large area of the exoplanet
($\pi R^{2}$) dominates the area of the atmospheric annulus ($ 2\pi
RNH$) in spite of the greater stellar intensity ($B^{*}_\lambda$).
Evaluating the ratio of these fluxes for $N=3$, $H=750$~km, $R=1.0
\times 10^{5}$~km, and for stellar and exoplanet temperatures of
$6000$~K and $1200$~K, respectively, we find that the ratio equals unity
for $\lambda = 2.6~\micron$.  A rigorous treatment would of course
include the line profiles and many other factors, but the inevitable
tendency is that, for wavelengths longward of the K photometric band,
occultation spectroscopy becomes increasingly favorable over
transmission spectroscopy in terms of the amplitude of the potential signal.

Occultation spectroscopy is essentially direct IR spectroscopy of the
exoplanet, and in principle this could be done at most orbital phases,
and also done for planets which do not transit. However, it requires
detection of the small-amplitude exoplanet spectrum (the precise
nature of which is unknown), superposed on a strong stellar
background.  The measurement becomes much easier if the signal is {\it
modulated predictably}.  The disappearance of the exoplanet
spectrum during secondary
eclipse is therefore an essential part of our technique. Note,
however, that modulation by variations in Doppler velocity could also be
used for all hot Jupiter systems, even ones without transits
\citep{wiedemann}.  Note also that the concept of occultation spectroscopy
is not new; it was successfully used to obtain the IR spectra of
Pluto and its moon Charon \citep{pluto}.

\section{OBSERVATIONS}

We observed two secondary eclipses, on UT 2001 July~8 and UT 2001 July~15, with
the ISAAC spectrometer on the VLT (Antu) at Cerro Paranal in
Chile.  The observations were performed in visitor mode, with
real-time decisions regarding nodding frequency, integration time,
etc.  Both nights were clear, with $10-20\%$ relative humidity, and
seeing in the range from $0.5-0.7$ arcsec.  The zenith column of
precipitable water vapor, as measured from telluric water lines in our
spectra, was $\sim6$ mm on July~8 and $1.6$ mm on July~15.

ISAAC is capable of imaging and spectroscopy in the wavelength region
between 1 and 5~\micron\ \citep{isaac}.  We used spectroscopic mode with
a resolving power of 3300, covering the 3.5 to 3.75~\micron\ 
range ($2667 - 2852$~cm$^{-1}$).  Both eclipses occurred within an hour
of transit across the local meridian, where the star reached a minimum
airmass of $1.38$. The spectra maintained good quality to the largest
airmass we observed ($2.51$).

At approximately half-hour intervals, we observed a comparison star,
\comp. The comparison star is within a degree of \tar,
with nearly the same B and V magnitudes and photometric colors (Table
\ref{stellar}). We observed both stars using a conventional `nod' technique,
placing them alternately at `a' and `b' positions on the slit, in the
order `abba', with $60$-second integrations at each slit position.  On
July~8 we obtained $36$ `abba' sets of \tar, and $20$ `abba' sets of
\comp, while on July~15 we recorded $41$ and $24$ sets, respectively.
About once per hour, we
obtained spectra of a continuum lamp for flat-fielding.  We did not
rely on the standard flats; instead, we recorded flat field exposures
immediately after each block of \tar\ observations without moving the
telescope or instrument, in order to avoid small but
non-negligible flexure effects.

\section{MODELED SPECTRUM}

We do not expect to achieve sufficient precision to detect the
planetary spectrum in each individual spectral resolution element;
instead, we look for the candidate signal by performing a
least-squares fit of the observations to a model `template'.  We
computed the template as the emergent flux from a model
atmosphere for the exoplanet, using a series of specific intensity
spectra at different values of $\mu$ (the cosine of the zenith angle).
These were computed from the
formal integral of the radiative transfer equation, using an LTE source
function; we neglected all
scattering terms, since the long-wavelength region is of interest and
we are assuming a cloud-free model.  Flux
was computed by a quadrature integration of the intensity spectra.
The spectrum synthesis was done at very high spectral resolution,
and convolved to the $0.8$ cm$^{-1}$ resolution of our ISAAC data using
a Gaussian instrument profile.  We adopted the parameters for the star
and planet given by \citet{cody}.

\begin{figure}
\plotone{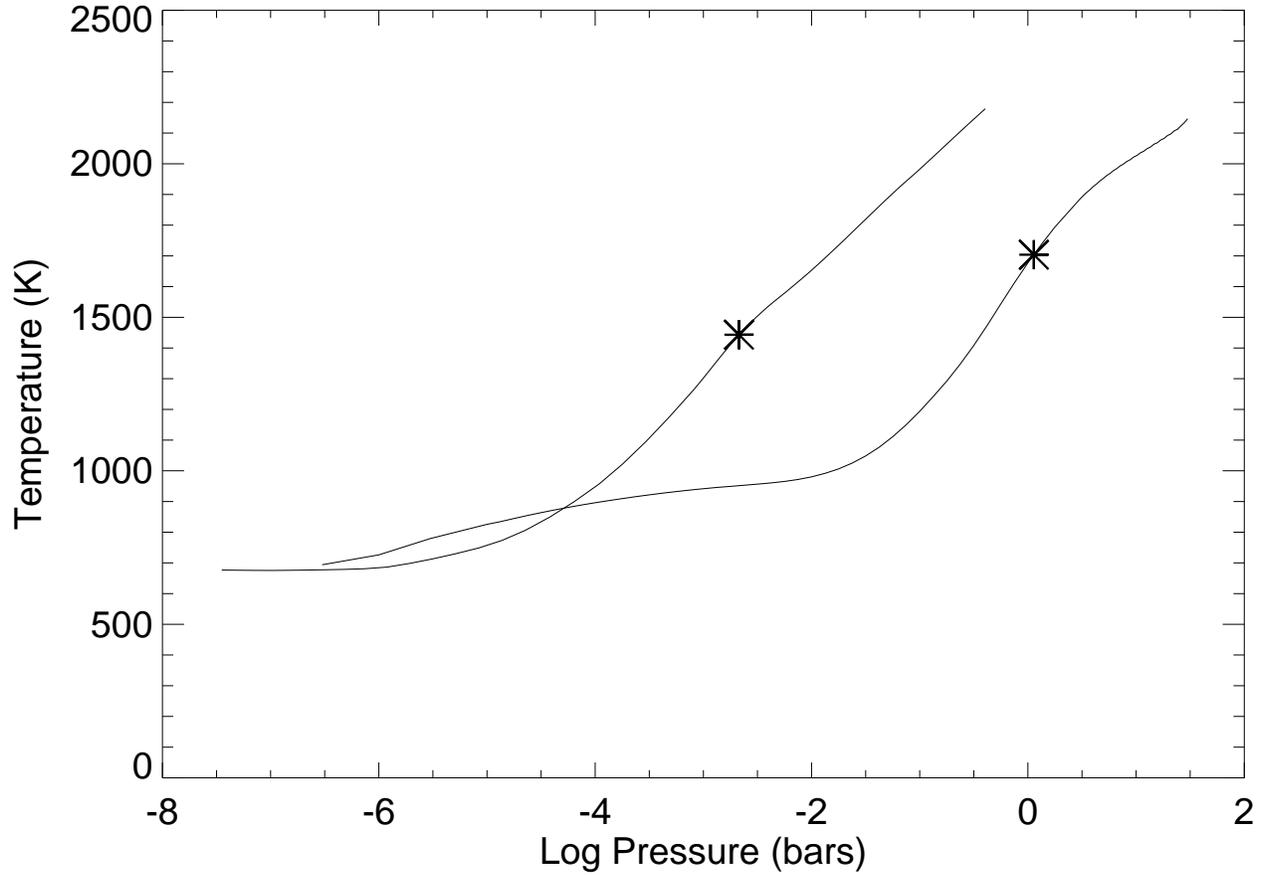}
\caption{Temperature vs. pressure structure for our fiducial model
for \tp\ (rightmost curve), and for a model by Seager
(2002). The asterisks mark the points of Rosseland optical depth unity
(Seager model) and $3.6\ \micron$ optical depth unity (fiducial
model).  \label{tp_prof}}
\end{figure}

Model atmospheres for the planet can vary widely, with one major
difference being the treatment of clouds and aerosols.  The maximum
thermal contrast in IR spectra is generally
obtained from clear atmospheres; one reason for this is the fact that
scattering processing in clouds tend to wash out strong absorption
features.  Moreover, the presence of clouds high in the atmosphere
effectively limits the detection of methane by allowing only the small
fraction of the column above the cloud to be probed by the observer. We have
therefore used the most optimistic case of the clear atmosphere
as a reference point.  Our fiducial model follows the
calculation of \citet{g00}, with the surface gravity appropriate
for \tp.  This model includes irradiation by
the star, but uses only gaseous opacities; condensates and
particulates are assumed to have settled out of the atmosphere forming
a deep cloud layer.  Figure \ref{tp_prof} compares the
pressure/temperature structure from our fiducial model with a recent
model from Sara Seager (2002, private communication).  The temperature
vs. optical depth structures of these models are similar, but the
fiducial model requires a much higher pressure (due to lower opacity)
to attain the same optical depth as the Seager model.  (We also
inspected a model for \tp\ kindly sent to us by
David Sudarsky (2002, private communication); it is slightly hotter than the
Seager model at all depths.)

We calculated the number densities of methane in the fiducial model,
using the simple analytic formulae given by \citet{bs99}.  The methane
mixing ratio peaks at $\sim 7 \times 10^{-6}$ at $25$~mbar pressure,
similar to the values for the 51 Peg model \citep{g00}. In
computing the template, we used the current best-available
experimental methane line parameters from \citet*{lb01}, now
incorporated into HITRAN \citep{hitran}, and the
calculation followed \citet{wiedemann} in other respects (partition
functions, etc.). Since line parameters are critical to
calculation of the template, we calculated spectra of hot methane
($773$ K) at ISAAC resolution for laboratory conditions, using both our
adopted line data and also theoretical line parameters from the
spherical top data system theory \citep{stds}.  We compared
these calculated spectra to laboratory measurements of hot methane at
$773$ K, measured by Steyert \& Reuter (2003, private communication) at
the same resolution.
Based on this comparison, we have
confidence in the HITRAN methane line data used to
compute our model
spectrum.  Since our spectral region also exhibits absorption due to
water, we experimented using water lines in our template, also based on
HITRAN.  Because the HITRAN database is incomplete at high temperatures,
we also experimented with using HITEMP, a HITRAN line
database extension for
higher temperatures, to obtain water line data.  This results in
abundant weak absorption lines
that do not significantly change the template.  We have therefore neglected
water absorption in the analysis until more complete information is
available for individual hot water lines in this spectral region.

We expressed the exoplanet flux spectrum as a ratio to the stellar
flux.  We computed the stellar flux from a gray atmosphere at
$T=5800$~K, reducing the adopted $6000$~K stellar temperature by $200$~K to
allow for the decrease in brightness temperature from the visible to
$3.6~\micron$, as documented for the solar spectrum by \citet{vernazza76}. 

\begin{figure}
\plotone{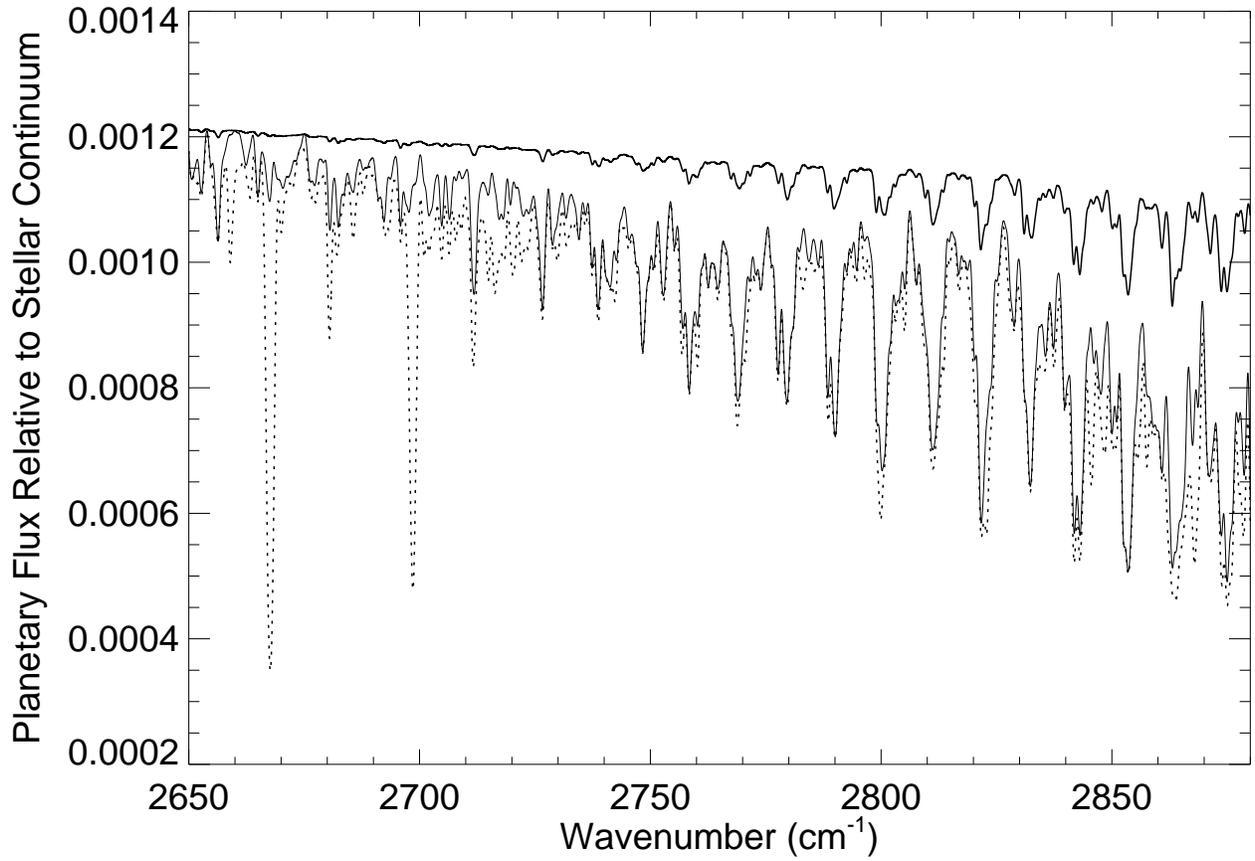}
\caption{Model template spectra containing only methane
lines (lower solid curve), and both methane and water lines (dotted
curve).  The upper solid curve represents the `methane only' template
with a factor of three increase in the continuous opacity, but
retaining the same temperature-optical depth relation. \label{models}}
\end{figure}

Figure~\ref{models} shows the modeled template spectrum, with and
without water lines from HITRAN included.  In the results quoted and
discussed below, we
have used the `methane only' version of the model template, for
reasons given above.  However, our results and associated errors do
not change significantly if `HITRAN water' or `HITEMP water' is
included in the model template.

We computed methane spectra from models intermediate between our
fiducial model and the Seager model.  If we increase the continuous
opacity in the fiducial model, while preserving the same
temperature-optical depth relation, the methane bands weaken
rapidly (again, see Figure~\ref{models}). The greater continuous
opacity pushes all line formation to
lower pressure layers, giving lower column densities `above the
continuum.'  The mixing ratio of methane decreases at low pressures
\citep{bs99}, further lowering the methane column density. 
Our fiducial model is a limiting case of an exceptionally clear
atmosphere, and many other models for the exoplanet will not
exhibit significant methane features, thereby escaping the test of our
present analysis.

\section{DATA ANALYSIS}

Analysis of the \tar\ observations took place in four stages: 1)
extraction of spectra from the 2-D data frames, 2) removal of the
telluric (and the few intrinsically stellar) absorption features
from the spectra, yielding residuals which potentially contain the
exoplanet spectrum, 3) estimation of the `amplitude' of the
exoplanet's spectrum in each residual spectrum, and 4) fitting of the
aggregate results to a secondary eclipse curve.  Since our present
(and future) results depend critically on the fidelity of our data
analysis, we describe this process in some detail, and we also
describe checks made to insure the integrity of the analysis.

\subsection{Spectral Extraction}

The spectral frames were first cleaned of `hot pixels' and temporary
anomalies such as energetic particle events.  Temporal
sequences of frames at
a single nod position were examined to locate hot pixels based on
their variability, and these pixels were given zero weight in
subsequent analysis.  Energetic particle events were identified and repaired
separately using a median filter applied to the temporal sequence.  The cleaned
`abba' set was combined by adding the `a' frames and subtracting the
`b' frames to produce a difference image, which was then divided by a
flat-field frame.  The flat-field frame was computed as an average of
the individual flats during the night. We verified that temporal
variations in the flats were negligible for purposes of our analysis.

At an excellent IR site such as Paranal, changes in the thermal
background are approximately linear with time during the several
minutes required to record an `abba' set.  Consequently, most of the
background emission is removed by construction of the difference image
a-b-b+a.  However, we found that a second order correction was
necessary to remove the background completely.  We compute the residual
background spectrum by summing the region between the a and b slit
positions in the difference image.  The relatively long slit used in
ISAAC facilitates the precision of this correction by providing a
$17$-arcsec length of background.  This residual background spectrum is
subtracted from each row of the two-dimensional difference image prior to the
extraction of the spectra.  Note that our technique differs from usual
IR practice, which would be to subtract the (negative) `b' spectrum from
the `a' spectrum after their extraction from the difference image,
which would automatically subtract the background to second order.
However, this would not maintain separate `a' and `b' spectra, which
we desired in order to verify that our results are independent of the
position of the spectra on the detector array.

We extracted the `a' and `b' spectra from the difference image using
the optimal extraction formalism of \citet{horne}.  The spatial profile
needed in the optimal extraction was constructed by fitting
polynomials in the intensity vs. wavelength distribution at each slit
position, and then sampling these curves in the spatial
direction at each wavelength.  In the process of extracting the
spectra, we noted that the approximate signal-to-noise ratio of each
spectrum ($\sim 200$) was close to the value expected based on the
noise from Poisson statistics of photoelectrons from the background.

\begin{figure}
\plotone{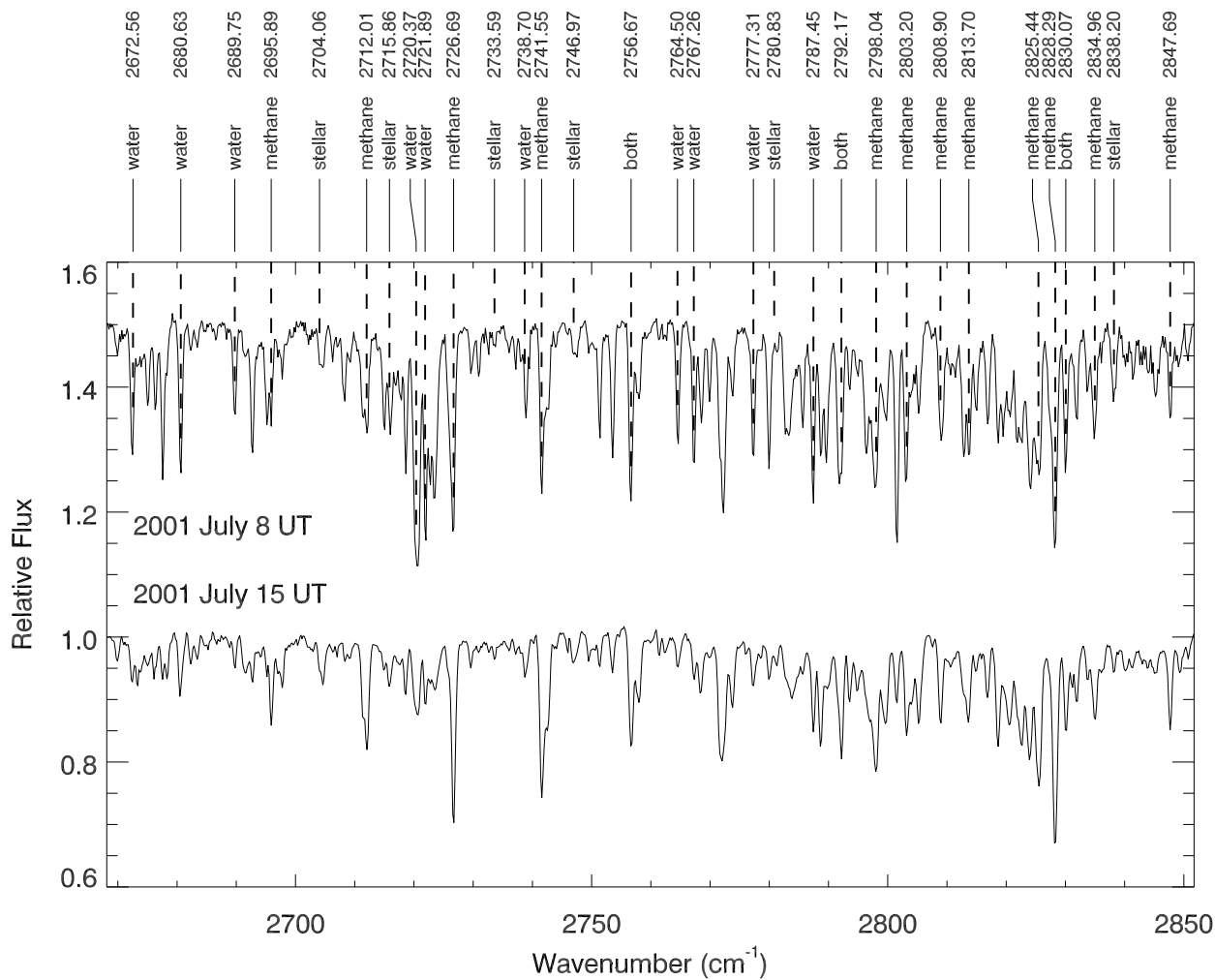}
\caption{Sample spectra of \tar\ for both observing nights. Most
of the line structure is telluric, due to methane and water vapor, as
noted by the line identifications.  A few stellar lines are also
marked \citep{atmos}.  Note that the telluric water vapor lines were
stronger on 2001 July~8.  The spectrum
from 2001 July~15 has been shifted by +0.5 for plotting purposes.
\label{spec}}
\end{figure}

At this stage of the analysis, the spectra were separated into sets
corresponding to July~8 or July~15, \tar\ or \comp, and `a' or
`b' slit positions.  Each of the 8 sets was analyzed independently of
the other sets. The individual spectra in each set were shifted
slightly using spline interpolation, so that the telluric absorption
features were all coincident. Frequencies in wavenumber for the
telluric lines were derived from high-resolution solar Fourier
transform spectra \citep{solar}, convolved down to the resolution of the ISAAC
data. (This accounted for any blending of telluric lines, which can
change their effective wavenumbers.) Wavenumbers for each point in our
ISAAC spectra were derived by spline interpolation using the telluric
lines as standards; the error of this procedure was $\sim
0.1$~cm$^{-1}$, much less than the $0.8$~cm$^{-1}$ spectral resolution.  A
single accurate wavenumber scale was thereby inferred for all the
spectra in each set.  Sample spectra from both nights are shown in
Figure~\ref{spec}.

\subsection{Telluric Correction}

The intensities in each spectrum fluctuate due to variable losses at
the ISAAC slit, as well as changes in the telluric line and continuum
absorption.  For each set of spectra, we made an intensity
normalization, followed by a correction to remove the telluric
absorption.  A pseudo-continuum was used as an aid to normalization;
it was determined as a spline fitted to a set of points in each
spectrum having the greatest intensity in pre-determined wavenumber
intervals.  We then used a wavelength-independent scale factor to
normalize each spectrum so that the integral intensity under the
pseudo-continuum was constant for all spectra.  That is, we enforce
the condition that the total intensity under each pseudo-continuum is
equal to the total intensity under the average pseudo-continuum; the
normalization factor $f_i$ is calculated from
\begin{equation}
\label{eq:factor}
f_i = \frac{\sum_{\lambda} \bar{c}}{\sum_{\lambda} c_i}
\end{equation}
where $\bar{c}$ represents the average pseudo-continuum, $c_i$
represents the fitted pseudo-continuum to spectrum $i$, and $f_i$ is
the factor by which each spectrum $i$ is multiplied.  Thus, the
normalized spectra are calculated by
\begin{equation}
\label{eq:norm}
n_i = f_i s_i
\end{equation}
where $n_i$ is the normalized spectrum and $s_i$ is the original spectrum.

In the normalized
spectra, at each wavelength element, we fit a least-squares line to
the natural log of intensity with airmass, $\ln n_i(x,\lambda)$.  The
fitted slope $b_{\lambda}$
was used to correct each intensity value to the minimum airmass in
that set of spectra (this corrects for airmass-dependent telluric
absorption).  That is, the airmass-corrected spectrum $n^\prime_i$ is
obtained from
\begin{equation}
\label{eq:airmass}
n^\prime_i = \exp \left( \ln n_i(x,\lambda) - b_{\lambda} (x -
x_{min}) \right)
\end{equation}
Each set of corrected spectra was then averaged.
Residual spectra were calculated by subtracting the average spectrum
and dividing the difference by a continuum fitted to the average
spectrum.  This produced residual spectra given by 
\begin{equation}
\label{eq:resid}
r_i = \frac{n^\prime_i - \bar{n^\prime}}{c_{\bar{n^\prime}}}
\end{equation}
which presumably contain the candidate exoplanet signal.

It is important to note several properties of the residual spectra.
Division by the pseudo-continuum of the average spectrum leaves the residuals
expressed in units of the stellar continuum.  However, since we divided by
only one continuum curve per data set, temporal changes in intensity
were not masked by fluctuating errors in continuum fits.  Also, the
subtraction of the average spectrum means that the sum of the
residuals at each wavelength is identically zero.  Thus, only variations
in the spectra survive this process. In particular, the exoplanet
spectrum will survive (since it varies due to the eclipse), but the
intrinsically stellar lines \citep{atmos} are subtracted.

\subsection{Higher-order Corrections}

Several additional corrections were needed before the residuals could
be compared with the model spectrum.  First, we noticed that a broad
absorption feature was evident in some of the residual spectra, and it
corresponded to a blend of water lines near $2720~\mbox{cm}^{-1}$.
The depth of this feature relative to the stellar continuum was $\sim
30\%$ greater on the more humid night (July~8), and variations of the
depth of the feature on either night were about $\pm 2\%$.
To correct these variations, we obtained a low-noise
residual water spectrum by differencing the natural logarithms of the
average spectrum for July~8 and July~15 (for both stars, and for both
the `a' and `b' spectra, separately).  This exploits the (fortuitous)
fact that the precipitable water vapor column was appreciably
different on the two nights, but telluric methane absorption was the
same.  (Unlike the concentration of water in the troposphere, which is
highly variable, methane is effectively constant because it is long-lived and
therefore well-mixed.)
We tested each residual spectrum to see whether a significant
correlation existed with the water vapor spectrum.  If it did, then we
removed it from the residuals by least-squares fitting.

\begin{figure}
\plotone{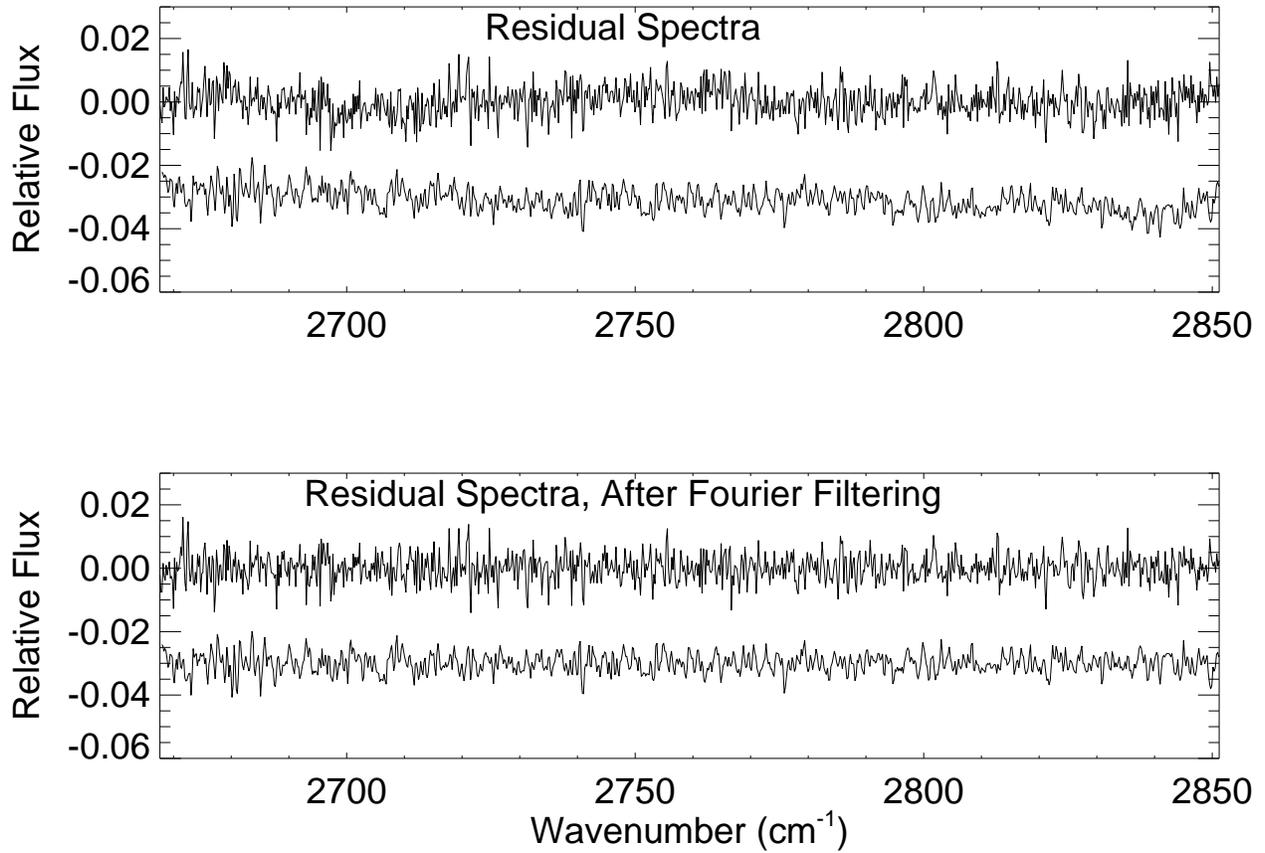}
\caption{Residual spectra corresponding to the example spectra shown
in Figure \ref{spec}.  Upper panel shows the residuals
from the extracted spectra after telluric correction.  Lower panel
shows the same residuals after the Fourier filtering process.  The
residual spectrum from July~15 (dashed line) has been shifted by -0.03
for plotting purposes.
\label{resid}}
\end{figure}

We also noticed that the residuals sometimes exhibit variations at low
spatial frequencies (see upper panel of Figure \ref{resid}). These
baseline effects
(typically $\sim 0.2-0.3\%$) are not surprising, because our analysis
treats each wavelength point independently.  (We do not introduce
artificial baselines - e.g., by continuum fitting - but neither do we
suppress real baseline variations.)  A Fourier analysis also revealed
a spike of excess power at the Nyquist frequency ($0.5$ cycles per pixel),
which is commonly seen in IR data, usually indicating a difference in
the two output amplifiers reading alternate columns of the detector
array.  We removed these effects by computing the Fourier transform
for each set of residuals; then we zeroed-out an appropriate number of
low frequency elements, as well as the DC offset (zero frequency) and
the Nyquist frequency components. The number of low frequencies
removed (up to $0.0079$ cycles per pixel) was taken to be the maximum
possible without impinging on frequencies where the exoplanet model
spectrum exhibited significant power.  The same filter operations were
performed on the model template spectrum as on the observed residuals.
Then we inverse-transformed the filtered residuals and template, which
were then suitable for comparison.  The effect of this Fourier
filtering on the observed residuals is seen the lower panel of Figure
\ref{resid}.

Even after the removal of variable water vapor absorption, and Fourier
filtering, the residuals needed an additional correction.  We
discovered that they contained an excess of `outliers', i.e., points
more than several standard deviations from zero.  We exploited our
comparison star observations to find those wavelengths which
tended to become outliers and reject them from the analysis. The
procedure was applied to the `a' and
`b' spectra separately, since we suspected low-amplitude `hot pixels'
as a likely cause of outliers. We applied a wavelength-dependent
`weighting mask' to the residuals, where the weight of the mask
(normally unity) was set to zero when a particular wavelength was
found to deviate from zero by more than $2.5\sigma$ in two or more
spectra for both \tar\ and \comp.  The requirement that points
are zero-weighted only if that wavelength tends to be discrepant in
spectra from
{\it both} stars assures that this procedure
does not bias the results.  About 7\% of the points were zero-weighted
using this method; we found that our final results were remarkably
insensitive to the limits for the rejection.  The reason for this is
that the total number of outliers was much less than $7\%$, because a
given point was always zero-weighted when it met the above criterion,
even if it was `well-behaved' in the majority of spectra.  The
standard deviation of the Fourier-filtered and masked residual
spectra was in the range $0.003$ to $0.006$.

\subsection{Fit to the Exoplanet Spectrum}

Prior to fitting the model template and Fourier filtering, the
residuals are corrected for the heliocentric Doppler shift due to the
relative motion of \tar\ with respect to the Sun \citep{nm02}, as
well as the geocentric Doppler shift due to the relative motion of the
Earth around the Sun.  We have also corrected for the non-negligible
Doppler shift due to the orbital motion of the exoplanet with
respect to \tar.  The residuals are shifted in wavenumber to the
rest frame of the exoplanet, and the model spectrum (already in the
exoplanet rest frame) is interpolated onto the same wavenumber grid.
Note that although the correction is small, we have corrected each
residual (each frame) for the Doppler effect, based on its time of
observation.  The residuals calculated for \comp\ are analyzed in an
identical manner.  Although \comp\ is not known to host a planet, we
apply the same Doppler correction as we did to \tar, but we calculate
the correction based on the time of observation of each \comp\ residual.

The comparison between each residual spectrum and the model exoplanet
spectrum is made using linear least-squares. For this purpose we use
the `methane only' version of the model (see Section 4), since the state 
of the spectroscopic line data gives us the most confidence in this
version. (Our overall results do not change with the `methane plus
water' version.)  The result of a least-squares regression of the
residuals {\it versus} the model spectrum gives a quantitative estimate of
the degree to which each residual spectrum `contains' the exoplanet
model spectrum.  We refer to these estimates as `model amplitudes',
since they represent the degree to which the model spectrum appears in
each residual spectrum. The least-squares fit also estimates the
random error in the model amplitudes, typically $1.5 - 2.5$.  In other
words, the signal-to-noise ratio for detecting the modeled exoplanet
spectrum in a single set of residuals is about $0.5$.  This is
sufficient precision to expect detection of the exoplanet spectrum in
the average of $\sim 50$ residual spectra, provided that the
signal-to-noise ratio increases as the square root of the number of spectra
averaged.

\subsection{Checks on the Analysis}

We have checked our analysis procedures in several ways.  Since we use
linear least-squares, the analysis should have the property that
averaging the best-fit model amplitudes for a set of residual spectra
should give the same result as fitting to the average of those
residual spectra.  We verified that our results satisfy this identity
exactly, {\it if} the time-variable Doppler shift applied to the
residuals is neglected (with Doppler shifts included, the equality is
approximate.)  We have also confirmed that our numerical procedures
do not attenuate potential signals.  We added a synthetic signal,
identical to the modeled spectrum, at the earliest practical stage of
the analysis (immediately after the extraction of the spectra from the
2-D data frames). The least-squares solutions recover this signal (as
a difference, out of eclipse minus in eclipse) with a best-fit model
amplitude of near unity.

\begin{figure}
\plotone{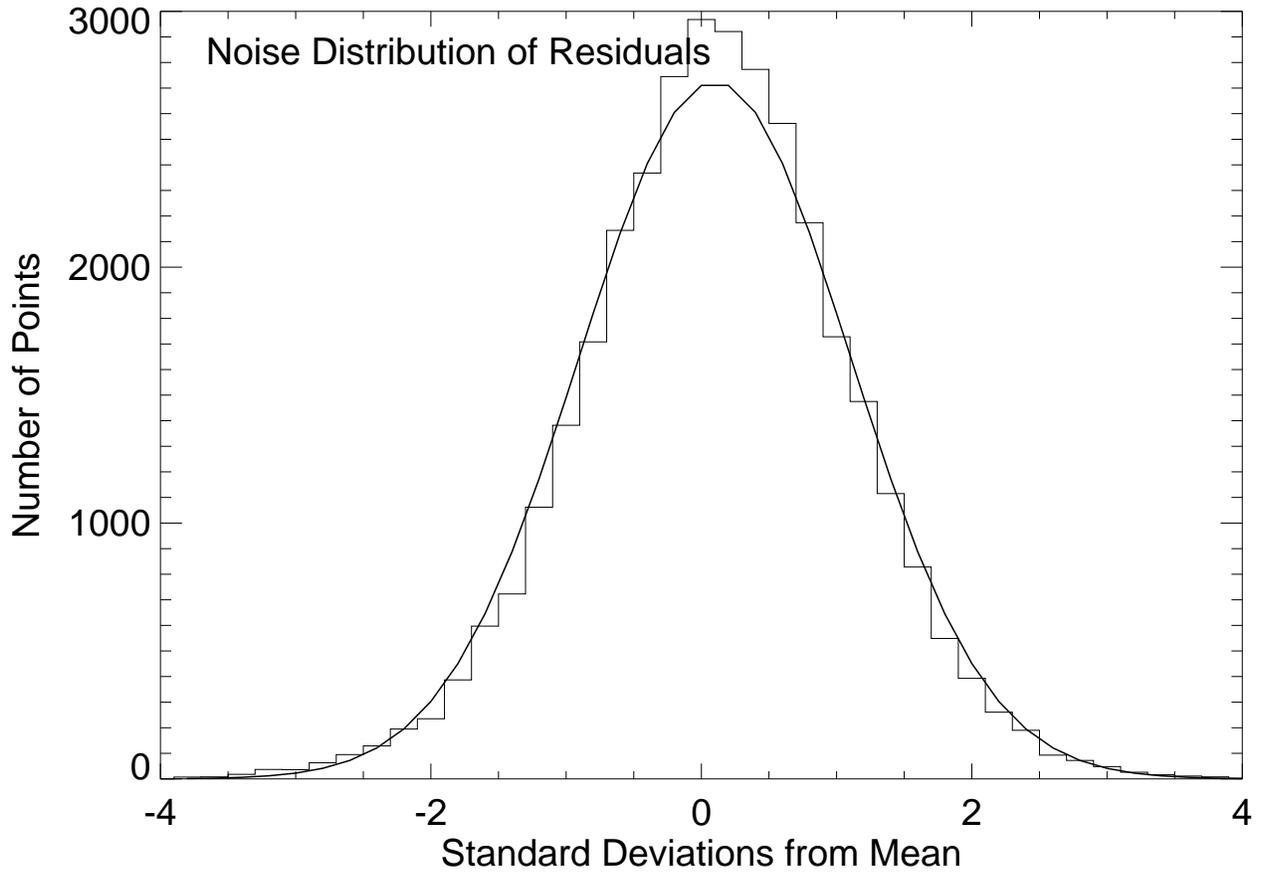}
\caption{Noise distribution of the residuals for the target `a'
spectra on the UT 2001 July~8.  Solid line indicates the theoretical value
based on the Gaussian probability distribution. \label{noise_r}}
\end{figure}

Another important check on our analysis is to examine the nature of
the noise, specifically the distribution of noise as compared to a
normal error distribution.  This is important because our error
estimates implicitly assume that the data reflect a normal error
distribution.  Figure \ref{noise_r} shows the error distribution for
all of the points (after masking) in the `a' residuals of \tar\ from
July~8.  Apart
from an excess of points within $1\sigma$ of zero, the distribution
closely approximates a normal error curve.  Similar distributions are
found for the other sets of data. 

\begin{figure}
\plotone{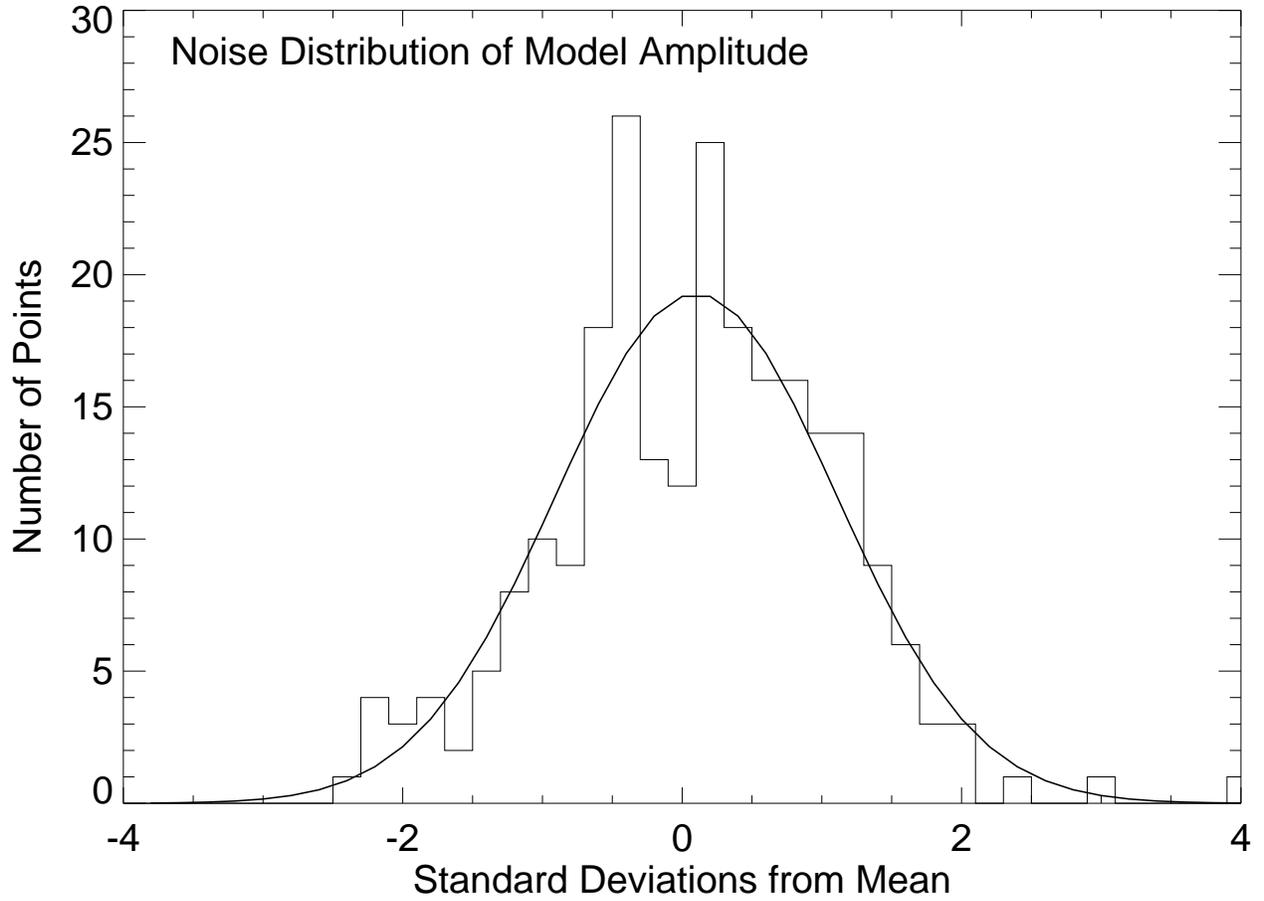}
\caption{Noise distribution for computed values of model amplitude.
All measurements (242) are included: both stars, both nights, and for both
the `a' and `b' spectra.  Solid line indicates the theoretical value
based on the Gaussian probability distribution.\label{noise_ma}}
\end{figure}

Figure \ref{noise_ma} shows the distribution of `model amplitudes.'
In this case we construct the distribution from the least-squares fits
to all residual spectra for both \tar\ and \comp\ from both nights; this
comprehensive inclusion is needed to provide sufficient points to
expect a reasonable approximation to the normal error curve.
Considering the number of residual spectra represented (242), the data are
in good agreement with the normal error curve.
As a further check on the errors, we computed the
variation in fitted model amplitudes which result from fitting
Gaussian random noise whose standard deviation varied from $0.003$ to
$0.006$, the same as the range of our filtered and masked residuals.  The
results of this simulation were in close agreement with the error
estimates derived from the least-squares fits to the actual residual
data.

\subsection{Secondary Eclipse Timing}

The timing of secondary eclipse depends on the orbital
eccentricity. If the eccentricity is identically zero (as might be
expected based on tidal circularization arguments) the secondary
eclipse will occur exactly mid-way between primary eclipses.  If the
eccentricity is non-zero, it can occur up to $\sim 90$ minutes earlier
or later (depending also on $\omega$).  Doppler observations to date
(G.~W.~Marcy 2002, private communication) give $e = 0.011 \pm 0.015$ with
$\omega = 156^\circ$, which would imply $\delta t = -31$~minutes,
i.e., a slightly earlier secondary eclipse.  We computed the time of secondary
eclipse as a function of the eccentricity, using $P=3.52474$~days
(G.~W.~Marcy 2002, private communication), and
the zero-point from primary eclipse at $T_{0}=2451659.93675$~HJD \citep{bc01},
and we corrected for light travel time.  We checked our calculations
against a primary eclipse ephemeris by David Charbonneau
(2002, \url{http://www.astro.caltech.edu/$\sim$dc/frames.html}), obtaining
essentially
identical results, and we did a similar check for the time difference
introduced by using a non-zero eccentricity (D.~Charbonneau 2002,
private communication).  Given a calculated time
for secondary eclipse, we fit a simple eclipse curve to our model
amplitude data by linear least-squares.  Our eclipse curve uses two
levels (in- and out-of-eclipse) connected by straight lines. The time
from first to fourth contact was taken to be $184.3$~minutes, and from
second to third contact $132.2$~minutes \citep{cb02}.  Note that, due to the
subtraction of an average spectrum in our data analysis, only the {\it
amplitude} of the eclipse curve is significant; the `zero-point' in
the fitted eclipse curve is not meaningful.

\section{RESULTS AND DISCUSSION}

Given the precision achieved by our analysis, we would expect to
detect the exoplanet spectrum if it is represented by the irradiated,
cloudless, low-albedo atmosphere, with thermochemical equilibrium
abundances of methane (and water), and also provided that the timing
of the secondary eclipse is known and the stellar and exoplanet
parameters (radii, etc.) are exactly as adopted. We first conclude
that this precision can indeed be achieved using ground-based
observations. The recent results and discussion of \citet{bc02} imply
that ground-based detection is feasible, since these authors came
within a factor of $\sim 3$ of the required precision using data from
only one night in poor weather.  Our experience, using a somewhat
different technique, confirms that the requisite precision can be
obtained if good observing conditions prevail.

\begin{figure}
\plotone{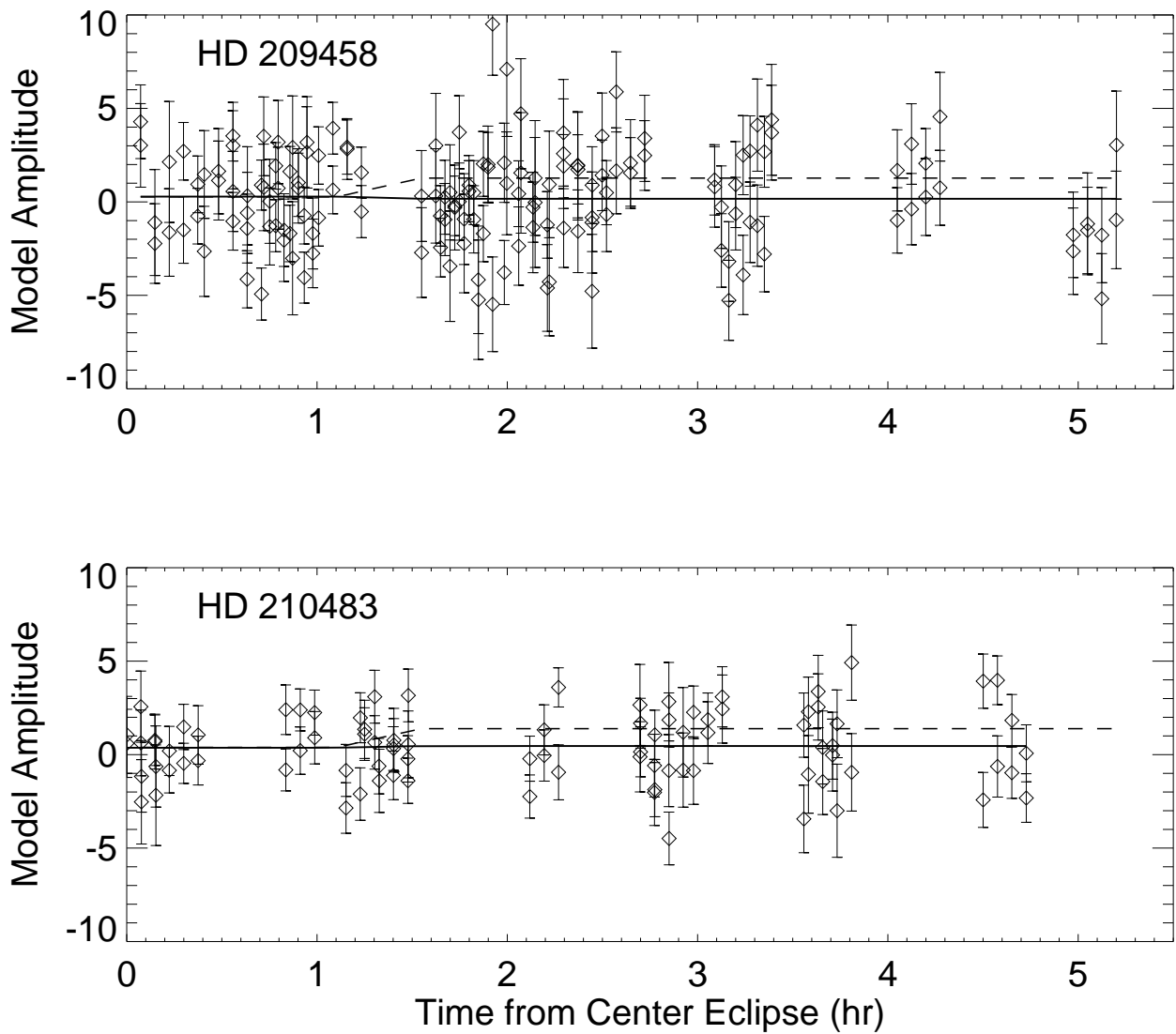}
\caption{Fitted secondary eclipse curves (solid line) for \tar\
(upper panel) and the comparison star (lower panel). These fits assume
the current best-fit Doppler value of the orbital eccentricity,
$e=0.01$, resulting in the eclipse being $31.1$ minutes earlier than
that of an orbit with zero eccentricity. The dashed lines represent an eclipse
amplitude of unity. \label{ma_e.01}}
\end{figure}

\begin{figure}
\plotone{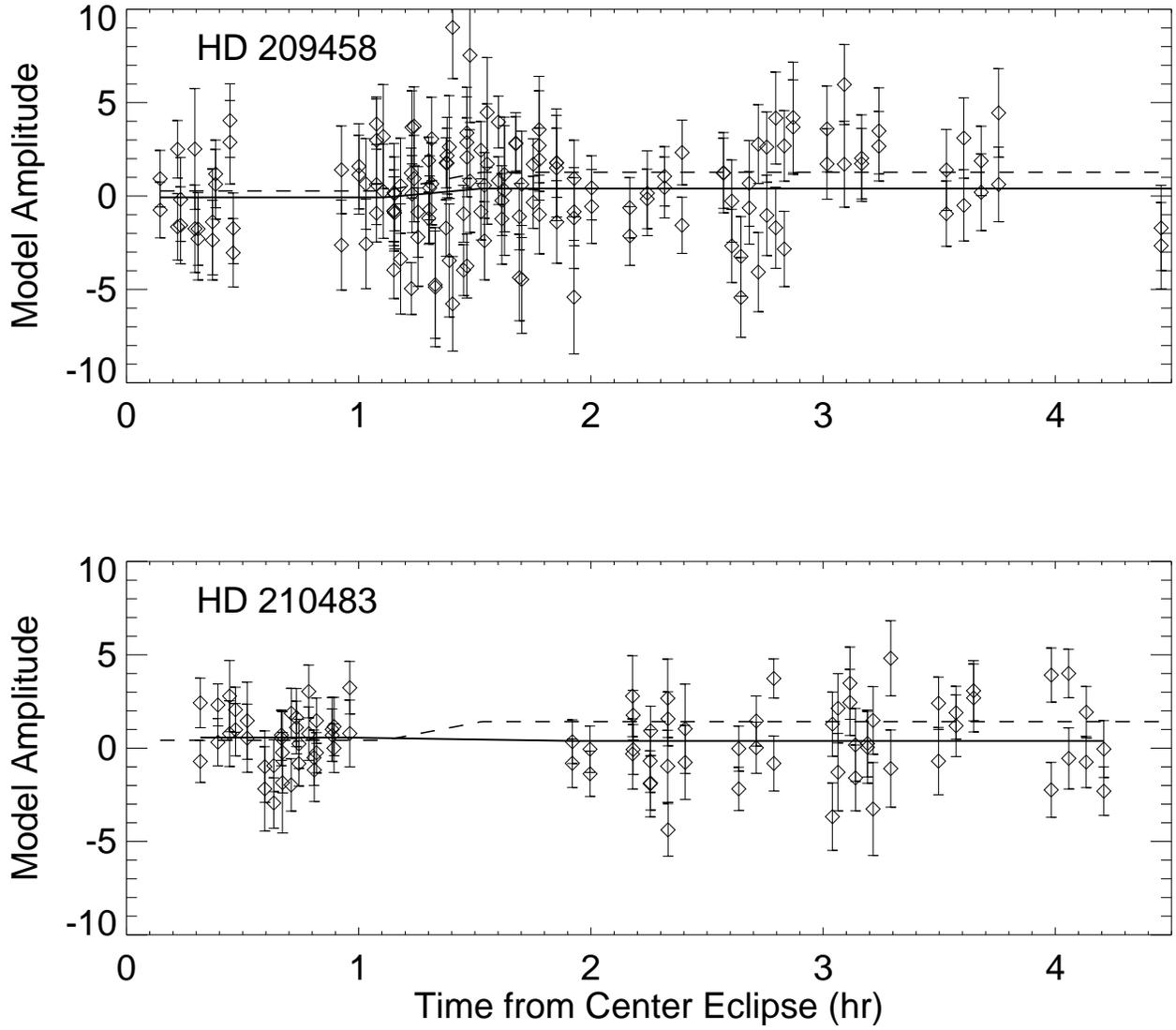}
\caption{Fitted secondary eclipse curves (solid line) for \tar\
(upper panel) and the comparison star (lower panel). These fits assume
that the orbital eccentricity is zero. The dashed lines represent an eclipse
amplitude of unity. \label{ma_e0}}
\end{figure}

Our results for the secondary eclipse are illustrated in
Figures \ref{ma_e.01} and \ref{ma_e0}, showing the
derived `model amplitudes' for each \tar\ spectrum (`a' and `b' for
both nights) {\it versus} time
from the assumed mid-point of secondary eclipse.  Recall that the
error bars for individual spectra were derived from the least-squares
fits to the residuals after telluric corrections, Fourier filtering,
and masking. We fit an eclipse curve to the aggregate results at each
assumed eclipse time. (We fold the curve about the mid-point of
secondary eclipse for the figures.)  Also, there is no evidence that
the results from the `a' and `b' spectra analyzed separately are
significantly different.  Figure \ref{ma_e.01} shows the
result for $e=0.011$, i.e., taking the slightly non-zero eccentricity
from the Doppler data at face value, again for all spectra.  The solid
line is the
least-squares estimate of the eclipse amplitude, $-0.1 \pm 0.3$, and
the dashed line shows the levels corresponding to an eclipse amplitude
of unity (i.e., if the model template spectrum correctly represents the
exoplanet, and the timing is correct.)  Figure \ref{ma_e0} shows
the corresponding result ($0.5 \pm 0.4$) under the assumption that the
eccentricity is identically zero.  In each case the reduced
chi-squared of the fit is $\sim 1.5$, indicating that the scatter of
the data is only slightly larger than the independently-assigned error
bars.  Note that each plot for \tar\ is accompanied by a similar
fit to the comparison star, shown in the lower panels.  In both cases,
the comparison star shows no significant change, as expected.

Adopting $e=0.011$ from the Doppler data, our results for non-detection
of the secondary eclipse ($-0.1 \pm 0.3$) exclude the model template
spectrum at the $> 3\sigma$ level. Given our error distributions, this
case is firmly rejected.  Assuming an eccentricity of zero,
non-detection of the secondary eclipse is ambiguous ($0.5
\pm 0.4$).  However, even if we explore eccentricities over the entire
plausible range from $0.0$ to $0.03$, and assume $\omega$ values in
the first and second quadrants, the fitted secondary eclipse
amplitudes never reach unity.  Instead, the amplitudes vary from
$-0.2$ to $+0.9$, and the amplitudes for the comparison star similarly range
from $-0.5$ to $+0.9$.  At a given eclipse time, the amplitudes for
the two stars are uncorrelated, as expected when sampling the noise
envelopes resulting from fitting independent random data to eclipse
curves of variable timings.  We therefore conclude that the secondary
eclipse is not detected using our `methane template.'  Given the
similarity between \tar\ and the comparison star in terms of the
noise envelope, and noting the nearly identical reduced chi-squared
values, we conclude that \tp\ probably does not exhibit
significant methane absorption features, in agreement with the models of
\citet{sudarsky03}.  Methane absorption in the combined light spectrum
(star plus planet) has a total equivalent width of no more than
$0.025$~cm$^{-1}$ within our absorption bandpass ($2667 - 2852$~cm$^{-1}$).

Deviation of the exoplanet atmosphere from our fiducial model is not
difficult to explain.  As discussed in Section 4, models of this
exoplanet must be exceptionally clear if significant methane
absorption is to appear in their spectrum.  In this respect our
fiducial model represents a limiting case.  Our results are certainly
consistent with the suggestion
of a cloudy atmosphere based on the observed low sodium abundance
\citep{cb02}. However, even in a clear atmosphere there may be other viable
explanations for both the sodium result \citep{barman02}, and
our present result (e.g., photochemical depletion of
methane). Nevertheless, it is significant that we can now begin to
limit the range of parameters of viable exoplanet models using {\it
ground-based} observations.

\section{Acknowledgements}

Jeremy Richardson is supported in part by a NASA Graduate Student Researchers
Program Fellowship, funded by the Office of Space Science at NASA
Headquarters (grant number NGT5-50273). Other aspects of this research
were supported by the NASA Origins of Solar Systems program.  We
extend special thanks to Gianni Marconi, the VLT night astronomer, who
contributed substantially to the success of these technically
challenging observations. We thank
Dave Charbonneau for several discussions concerning the secondary
eclipse, and for sharing his insights generally.  We benefitted from
discussions with Sara Seager on the characteristics of exoplanet model
atmospheres, and we thank her for providing her recent
temperature/pressure relation.  We are grateful to Geoff Marcy for
communicating the results of the latest solution for orbital
eccentricity, and to Dave Sudarsky for sending us his
temperature/pressure profile for \tp. We also thank Linda Brown
for discussions on the status of methane line data.  Finally, we thank
the referee for helpful comments and suggestions.

\clearpage

\bibliographystyle{apj}
\bibliography{ms}

\clearpage

\begin{deluxetable}{crr}
\tabletypesize{\scriptsize}
\tablecaption{Stellar Photometric Data. \label{stellar}}
\tablewidth{0pt}
\tablehead{
\colhead{Parameter} & \colhead{\tar} & \colhead{\comp}
}

\startdata
V\tablenotemark{1} &7.648 &7.586\\
B-V\tablenotemark{1} &0.594 &0.585\\
b-y\tablenotemark{2} &0.361 &0.391\\
m1\tablenotemark{2} &0.174 &0.175\\
c1\tablenotemark{2} &0.362 &0.354\\
\enddata

\tablenotetext{1}{Hipparcos Catalogue \citep{perryman}}
\tablenotetext{2}{$uvby\beta$ Photometric Catalogue \citep{hauck}}
\end{deluxetable}

\clearpage

\end{document}